\documentclass[pra,showpacs,superscriptaddress,twocolumn]{revtex4}

\usepackage{epsfig}
\usepackage{graphicx}
\usepackage{amssymb}

\begin{document}

\title{Solvable model of a strongly-driven micromaser}

\author{P. Lougovski}
\affiliation{Max-Planck Institut f\"{u}r Quantenoptik,
Hans-Kopfermann Strasse 1, 85748 Garching, Germany}

\author{F. Casagrande}
\affiliation{INFM - Dipartimento di Fisica, Universit\`{a} di Milano,
Via Celoria 16, 20133 Milano, Italy}

\author{A. Lulli}
\affiliation{INFM - Dipartimento di Fisica, Universit\`{a} di Milano,
Via Celoria 16, 20133 Milano, Italy}

\author{B.-G. Englert}
\affiliation{Department of Physics, National University of Singapore,
Singapore 117\,542, Singapore}
\affiliation{Max-Planck Institut f\"{u}r Quantenoptik,
Hans-Kopfermann Strasse 1, 85748 Garching, Germany}

\author{E. Solano}
\affiliation{Max-Planck Institut f\"{u}r Quantenoptik,
Hans-Kopfermann Strasse 1, 85748 Garching, Germany}
\affiliation{Secci\'{o}n F\'{\i}sica, Dpto. de Ciencias, Pontificia
Universidad Cat\'{o}lica del Per\'{u}, Apartado 1761 Lima, Peru}

\author{H. Walther}
\affiliation{Max-Planck Institut f\"{u}r Quantenoptik,
Hans-Kopfermann Strasse 1, 85748 Garching, Germany}

\date{\today}

\begin{abstract}
We study the dynamics of a micromaser where the pumping atoms are
strongly driven by a resonant classical field during their transit
through the cavity mode. We derive a master equation for this
strongly-driven micromaser, involving the contributions of the
unitary atom-field interactions and the dissipative effects of a
thermal bath. We find analytical solutions for the temporal
evolution and the steady-state of this system by means of
phase-space techniques, providing an unusual solvable model of an open
quantum system, including pumping and decoherence. We derive
closed expressions for all relevant expectation values, describing
the statistics of the cavity field and the detected atomic levels.
The transient regime shows the build-up of mixtures of mesoscopic
fields evolving towards a superpoissonian steady-state field that,
nevertheless, yields atomic correlations that exhibit stronger
nonclassical features than the conventional micromaser.
\end{abstract}

\pacs{42.50.Pq, 03.65.Yz, 32.80.Qk}

\maketitle

\section{Introduction}

The micromaser~\cite{m.m.} is a fundamental system in cavity
quantum electrodynamics (CQED)~\cite{qed} where a single cavity
mode is pumped by a poissonian beam of excited two-level atoms, in
such a way that at most one atom is inside the cavity. While each
atom traverses the cavity, the entangled atom-cavity system
undergoes the Jaynes-Cummings (JC) interaction~\cite{JC} and, when
this is not the case, the field decays due to its coupling to
the environment. It has been demonstrated that the micromaser
exhibits interesting nonclassical features in the field statistics
and in the atomic correlations, like subpoissonian photon
statistics~\cite{subp}, trapping states~\cite{TS} and number
states~\cite{Fock} of the cavity field, and antibunched atomic
correlations~\cite{Englert1}.

Recently, it was shown that the effective coupling between an atom
and a single cavity mode can be drastically modified in the
presence of a strong external driving field~\cite{SAW}. For
example, under resonant conditions among the atomic transition,
the cavity mode and the external field, it is possible to engineer
a resonant JC and anti-JC interaction simultaneously. In this
case, the usual atom-field Rabi oscillations do not play any
further role, leaving their place to conditional field
displacements depending on the atomic internal states:
``Schr\"odinger cat states.'' Generation of entanglement and its
measurement is a very active field in CQED with diverse
implications in fundamental and applied quantum
physics~\cite{Englert1,Haroche}. From that point of view, a more
realistic study of this new scheme and its different variants,
where dissipative effects are fully taken into account, is
desirable.

In this paper we develop the theory of a strongly-driven
micromaser (SDM) and study the new features of the field
statistics and the atomic correlations. We formulate a master
equation for the cavity field density operator where the gain
originates in the coherent interaction of the field with two-level
atoms which are strongly driven by an external driving while they
are inside the cavity~\cite{SAW}. The losses in the master
equation stem from the interaction of the cavity field with a
thermal bath. The additional strong driving acting on the atoms
changes drastically the SDM dynamics as compared with the
conventional micromaser.

We show that the SDM master equation can
be solved analytically by means of phase-space methods, providing
an unusual solvable model of an open quantum system under realistic
conditions. In this way we are able to trace the temporal
evolution of the SDM field from an arbitrary initial state to the
final steady-state, which happens to be superpoissonian. We derive
closed expressions for the main quantities which characterize the
statistics of the cavity photons and the detected atoms. We find
that, despite the classicality of the SDM field steady-state, the
atomic correlations exhibit stronger nonclassical features when
compared with the conventional micromaser. The description of
system dynamics is illustrated also by numerical results which
support and complement the analytical ones. It is worth noting
that the present SDM system differs markedly from the recently
investigated coherently-driven micromaser~\cite{CL}, where an
external field continuously drives the cavity mode and no strong
driving regime is required.

In Sec.~\ref{sec2}, we present the atom-field Hamiltonian of the SDM
and illustrate its main dynamical features.
In particular, we show how the cavity field changes as a consequence of the
passage of one, two, or more atoms.
We pay careful attention to the crucial difference between the situations in
which the atoms remain unobserved and in which their final state is detected.
Then, in Sec.~\ref{sec3}, we
formulate the SDM master equation and derive its time-dependent
analytical solution.
As an application, we use it to calculate the expectation values of
various important observables of the photon field,
and of the atom counting statistics.
Numerical results are discussed in Sec.~\ref{sec4},
and conclusions are drawn in Sec.~\ref{sec5}.

\section{Hamiltonian and coherent dynamics}
\label{sec2}

The fully resonant interaction between one mode of a high-Q cavity
and a two-level atom strongly driven by a classical external field
can be described by the following Hamiltonian~\cite{SAW}
\begin{equation}
\label{H} H = \frac{\hbar g}{2}(\sigma^\dagger +
\sigma)(a^\dagger+a),
\end{equation}
where $g$ is the atom-cavity mode coupling constant, $a$
($a^\dagger$) the field annihilation (creation) operator, and
$\sigma = |g\rangle \langle e|$ ($\sigma^\dagger = |e\rangle
\langle g|$) the atomic lowering (raising) operator. The
Hamiltonian of Eq.~(\ref{H}), written in the interaction picture,
was derived in~\cite{SAW} in the strong driving regime $\Omega \gg
g$, where $\Omega$ is the Rabi frequency associated with the
external field.

A first feature of Hamiltonian in Eq.~(\ref{H}) is the presence of
resonant terms of both the Jaynes-Cummings type, $\sigma^\dagger a
+ \sigma a^\dagger$, and of the anti-JC type, $\sigma^\dagger
a^\dagger  + \sigma a$. Actually, the latter terms are usually
negligible in CQED, whereas they can be of importance in other
systems, like ion traps~\cite{ion}. A second feature, that is
discussed below, is the direct generation of the so-called
Schr\"{o}dinger cat states of the cavity field.

The Hamiltonian of Eq.~(\ref{H}) generates a time evolution that
is described by the unitary operator
\begin{equation}
\label{U} U(\xi) = D(\xi)|+\rangle \langle +| + D(-\xi)|-\rangle
\langle -|\,,
\end{equation}
where the parameter $\xi = -ig\tau/2$ is proportional to the
interaction time $\tau$;
$D(\alpha) = \exp{(\alpha a^\dagger - \alpha^*a)}$ is the unitary
displacement operator; and $|\pm\rangle =
(1/\sqrt{2}) (|g\rangle \pm |e\rangle)$ are the eigenstates of the
atomic operator $\sigma_{x}\equiv \sigma^\dagger + \sigma$ with
eigenvalues $\pm 1$, respectively. Starting from an
initial state with the cavity field in a state $\rho_{F,0}$ and
one atom injected in the excited state,
\begin{equation}
\label{ro0} \rho_{0} = \rho_{F,0}\otimes |e\rangle \langle e|\,,
\end{equation}
the density operator evolves to the state
\begin{equation}
\label{ro1} \rho_{1} = U(\xi)\rho_{0}U(-\xi)\,.
\end{equation}
We consider two cases, one in which the atom is not observed when
it leaves the cavity and the other in which it is detected in the
upper or lower level.

\subsection{Atoms are not observed}\label{sec2a}

If we are interested in the cavity field after the atomic transit
without measuring the state of the outgoing atom, the cavity field
density operator is
\begin{equation}
\label{roF1} \rho_{F,1} = \mathrm{Tr}_{A} \rho_{1} = \frac{1}{2}\left[
D(\xi)\rho_{F,0}D(-\xi)+ D(-\xi)\rho_{F,0}D(\xi) \right]\,,
\end{equation}
where $\mathrm{Tr}_{A}$ is the partial trace over the atomic degrees of
freedom, and we used Eqs.~(\ref{U})-(\ref{ro1}).
If the cavity is
initially in the vacuum state, $\rho_{F,0} = |0\rangle \langle 0|$,
its state turns into
\begin{equation}
\label{roF,1 bis}\rho_{F,1} = \frac{1}{2} \left( |\xi\rangle
\langle\xi| + |-\xi\rangle \langle -\xi| \right)\,,
\end{equation}
which is a statistical mixture of two coherent states with the
same amplitude and opposite phases.

It is of interest for what follows to describe the time evolution
of the cavity field in phase space. While for the vacuum state the
Wigner function is the Gaussian $W_{0}(\alpha) =
2\exp{(-2|\alpha|^2)}$, for the state of Eq.~(\ref{roF,1 bis}) we
have
\begin{equation}
\label{W1} W_{1}(\alpha)  = 2 \exp{
\left[ -2(|\alpha|^2 + |\xi|^2) \right] }\cosh(4|\xi|\mathrm{Im}\,\alpha)\,.
\end{equation}
This last expression shows that the interaction with a driven atom
can affect the rotational symmetry of the initial vacuum state.
The section of $W_{1}(\alpha)$ along the real axis remains a
Gaussian with the initial minimum uncertainty, while the section
along the imaginary axis is broadened, showing a two-peaked
structure that results from the superposition of two Gaussians
centered at $\alpha = \pm\xi$. The two peaks are well resolved if
$|\xi| \gg 1/2$, which requires that the system is operated in the
strong-coupling regime $(g\tau \gg 1)$.

\begin{figure}[!t]
\centerline{\includegraphics[width=68mm]{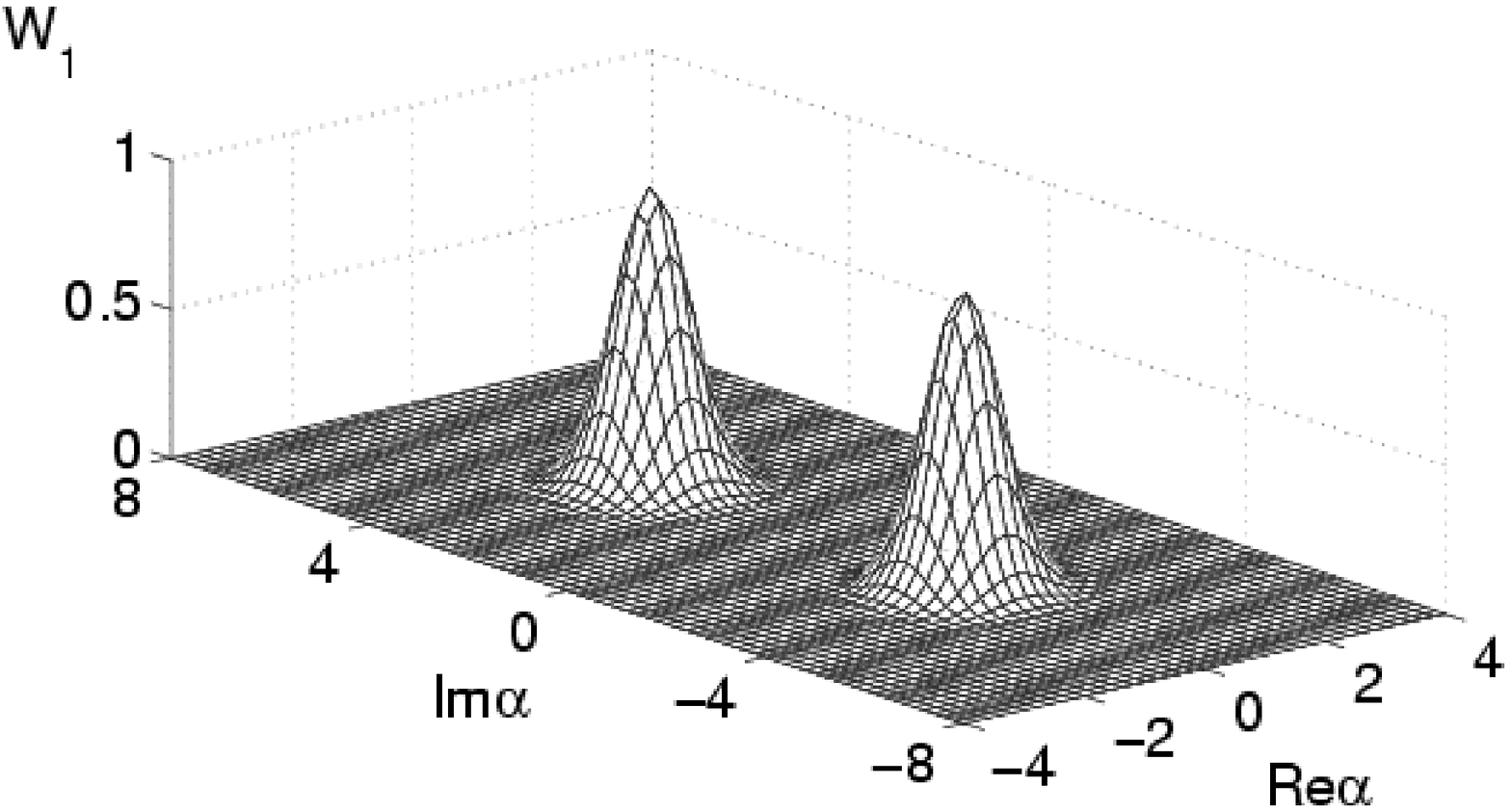}}
\centerline{\includegraphics[width=68mm]{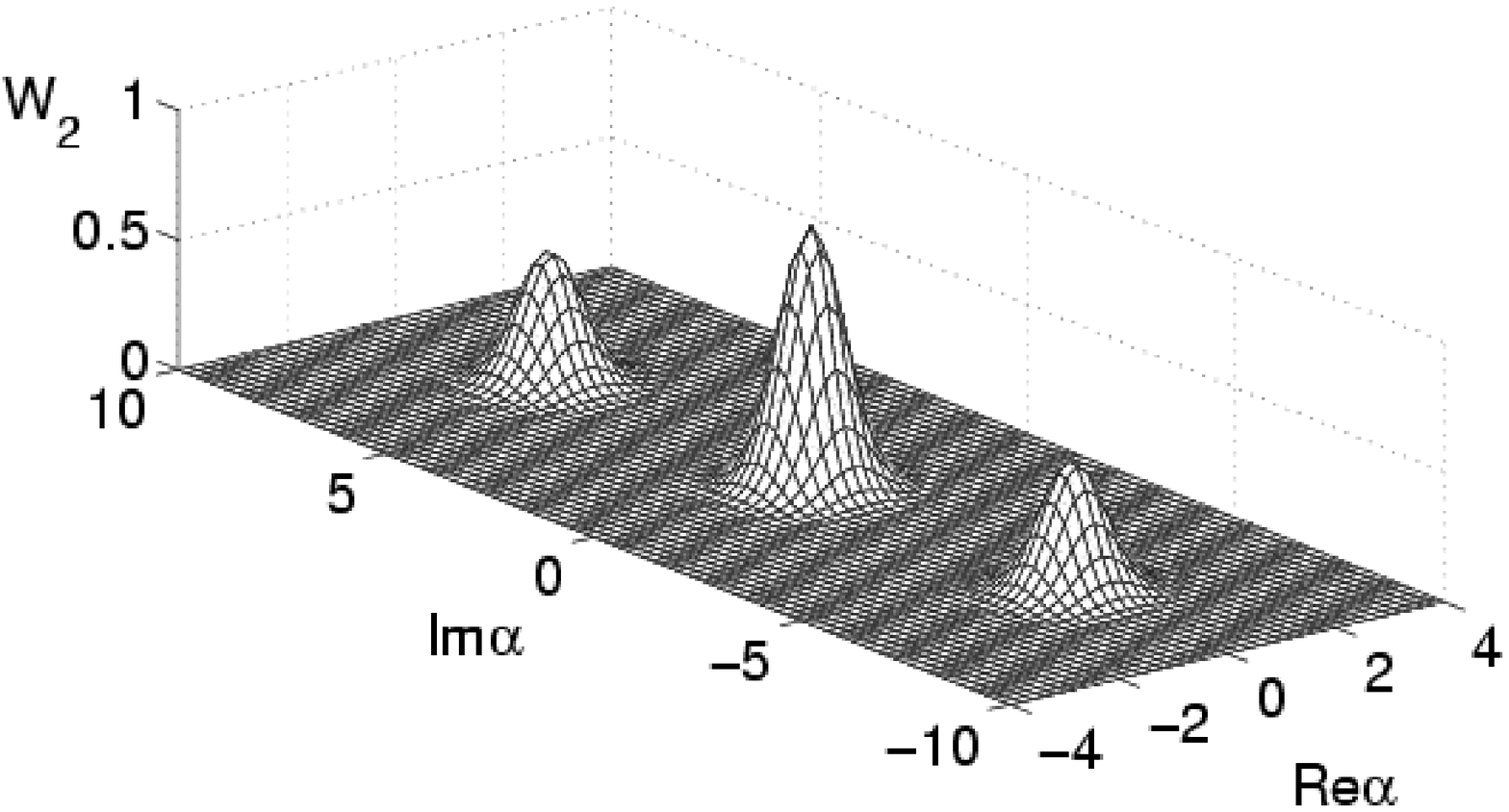}}
\caption{\label{fig1} Wigner distributions $W_1(\alpha)$ and
$W_2(\alpha)$ of the cavity field, see Eq.~(\ref{Wm}), in the case
of one and two crossing atoms, respectively, provided that no attempt
is made to determine the final internal
state of the atoms. Here,
as well as in all the following figures, $| \xi | = \pi$, the
cavity field is initially in the vacuum state, and the atoms in
the excited state.}
\end{figure}

If the cavity mode, initially in the vacuum state, interacts with
$1,2,\ldots ,m$ strongly driven atoms, all assumed to pass through the
cavity (one by one) with the same interaction time $\tau$ and to
leave it unobserved, the final field state is given by the
density operator
\begin{equation}
\label{roFm} \rho_{F,m} = \frac{1}{2^m}\sum_{n=0}^{m} \left(
\begin{array}{c}
  m \\ n
\end{array} \right)
|(m-2n)\xi\rangle \langle(m-2n)\xi|\,,
\end{equation}
which describes a mixture of $m+1$ coherent states. This state is
represented in phase space by the Wigner function
\begin{equation}
\label{Wm} W_{m}(\alpha) = \frac{1}{{2^{m-1}}}\sum_{n=0}^{m}
\left(
\begin{array}{c}
m \\ n
\end{array}
\right) \exp{ \left[ -2|\alpha - (m-2n)\xi|^2 \right]
}\,.
\end{equation}

Hence, in the strong coupling regime and with negligible
dissipation effects, multi-peaked cavity field distributions can
be generated in phase-space, where the number of peaks increases
with the number of driven atoms injected in the cavity. The
subsequent generation of such cavity field states is illustrated
in Fig.~\ref{fig1}, where we show the Wigner functions
$W_{1}(\alpha),W_{2}(\alpha)$ calculated from Eq.~(\ref{Wm}).
After the transit of m atoms the function $W_{m}(\alpha)$ exhibits
$m+1$ peaks, all centered on the imaginary axis and with a
center-to-center distance of $2|\xi| = g\tau$. If $m$ is even, the
peaks are centered at $\mathrm{Re} \alpha = 0$ and
$\mathrm{Im} \alpha = 0, \pm
2|\xi|,\ldots,$ where the central peak in the origin is the highest
one. If $m$ is odd, the centers are at $\mathrm{Re}\alpha = 0$ and
$\mathrm{Im}\alpha
= \pm |\xi|,\pm 3|\xi|,\ldots$, the highest peaks being at $\alpha =
\pm\xi$.

\subsection{Atoms are detected}\label{sec2b}

Now we consider the cavity field in the case in which the strongly
driven atoms are detected when they leave the cavity, e.g. their
state is determined by selective field ionization, where for
simplicity we assume perfect detector efficiency. We are again
interested in the cavity field properties after
interaction. Proceeding again from the initial condition of
Eq.~(\ref{ro0}), and using Eqs.~(\ref{U})-(\ref{ro1}), the cavity
field density operators after detecting the atom in the excited
state ${\rho}^{(e)}_{F,1}$ or in the ground state
${\rho}^{(g)}_{F,1}$ are
\begin{eqnarray}
{\rho}^{(e)}_{F,1}&=&\frac {\left[D(\xi)+ D(-\xi)\right]
\rho_{F,0}\left[D(\xi)+ D(-\xi)\right]}{2\left[
\mathrm{Re}\chi(2\xi) + 1 \right]}\,,\nonumber\\[1ex]
{\rho}^{(g)}_{F,1}&=&\frac {\left[D(\xi)- D(-\xi)\right]
\rho_{F,0}\left[D(\xi)-D(-\xi)\right]}{2\left[
\mathrm{Re}\chi(2\xi) - 1 \right]}\,, \label{ro1eg}
\end{eqnarray}
where $ \chi(\beta) = \mathrm{Tr}_{F}\left\{ \rho_{F,0} D(\beta)
\right\}$ is the characteristic function for symmetrical ordering
of the field operators~\cite{CG} and $\mathrm{Tr}_{F}$ is the
partial trace over the field variables. If the cavity is initially
in the vacuum state, then
\begin{eqnarray}
\label{ro1egbis} && {\rho}^{(e)}_{F,1} = \frac { |\xi\rangle
\langle \xi| + |-\xi\rangle \langle -\xi|+ |\xi\rangle \langle
-\xi| + |-\xi\rangle \langle \xi| }{2\left[ 1 + \exp(-2|\xi|^{2})
\right]} \,,
\nonumber \\
&&{\rho}^{(g)}_{F,1} = \frac { |\xi\rangle \langle \xi| +
|-\xi\rangle \langle -\xi| - |\xi\rangle \langle -\xi| -
|-\xi\rangle \langle \xi| }{2\left[ 1 - \exp(-2|\xi|^{2}) \right]}\,, \nonumber \\
\end{eqnarray}
which are pure states of the cavity field instead of the
statistical mixtures of Eq.~(\ref{roF,1 bis}). Actually, the
cavity field state vectors are $ {|{\psi}_{F,1}\rangle}^{(e),(g)}
\propto(|\xi\rangle \pm (|-\xi\rangle)$, that is the superposition of
two coherent states of the kind generated and monitored in the
dispersive regime of cavity QED in~\cite{cats}. More elaborated
superposition states were investigated in Ref.~\cite{SAW}, where
also many-atom states were considered. The mean photon number of
the field states of Eq.~(\ref{ro1egbis}) are
\begin{equation}
\label{N1} {\langle N_1\rangle}^{(e),(g)} = {|\xi|^2}\frac {1\mp
\exp(-2|\xi|^2)}{1\pm \exp(-2|\xi|^2) }\,,
\end{equation}
whereas $ \langle N_1\rangle = |\xi|^2 $ in the case of an
unmeasured atom of Sec.~\ref{sec2a}.

\begin{figure}[!t]
{\centering
\makebox[80mm][l]{(a)}\\[-3ex] \includegraphics[width=68mm]{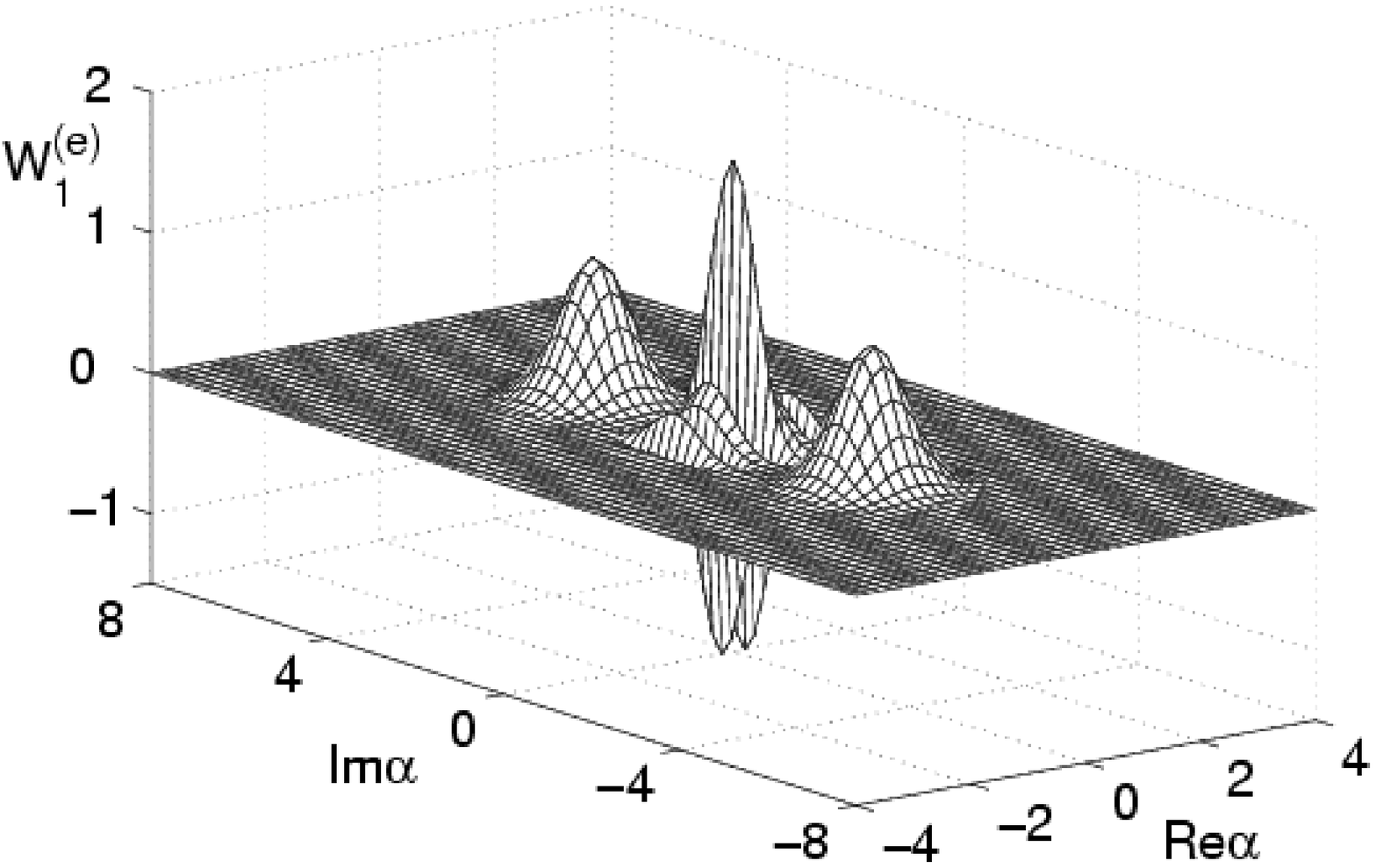}\\
\makebox[80mm][l]{(b)}\\[-3ex] \includegraphics[width=68mm]{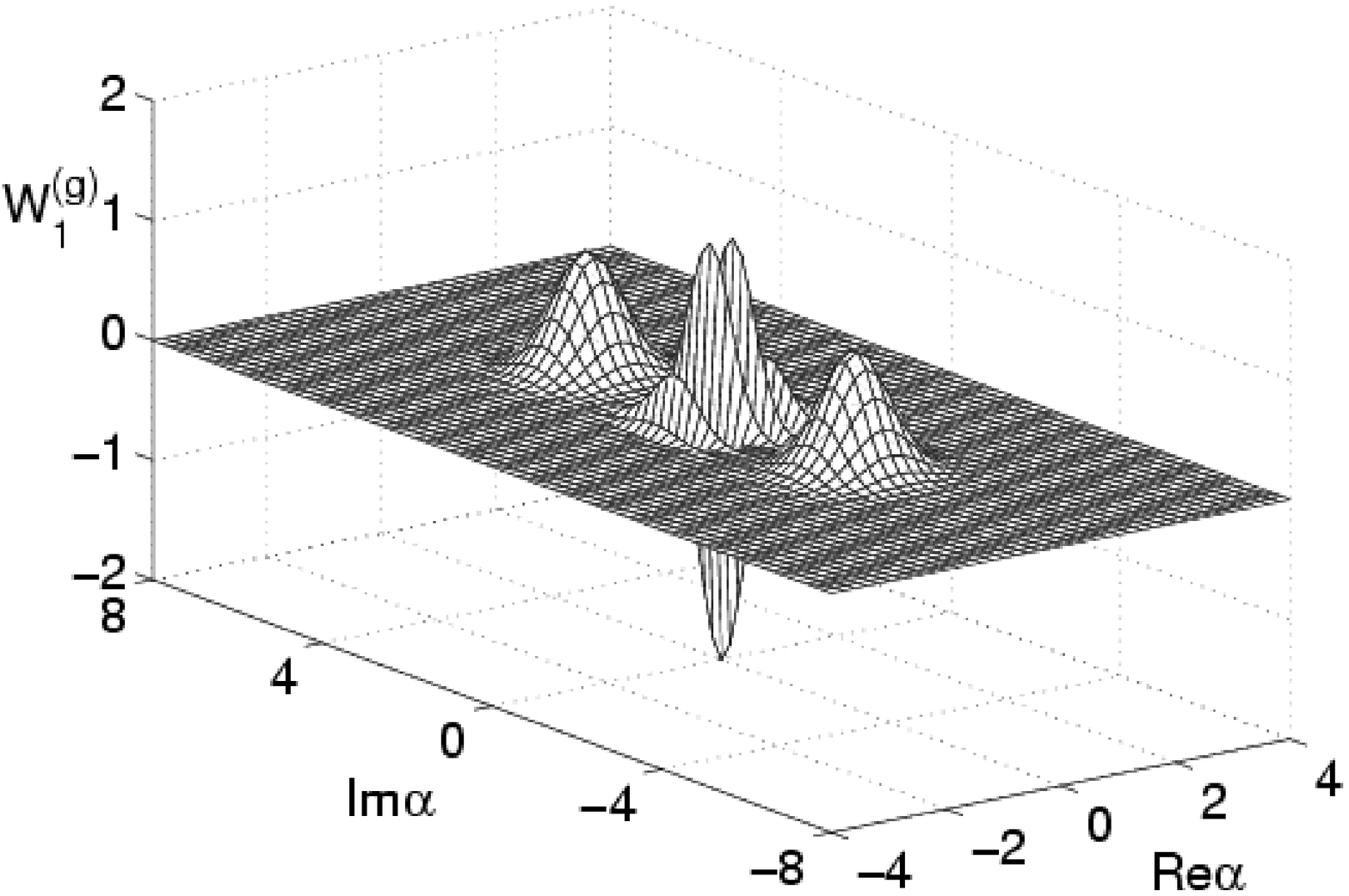}}
\caption{\label{fig2} Wigner distribution of the cavity field for
the case of one crossing atom detected (a)~in the excited state,
$W_1^{(e)} (\alpha)$; (b)~in the ground state, $W_1^{(g)}
(\alpha)$; see Eq.~(\ref{W1eg}).}
\end{figure}

The Wigner functions representing the states of
Eq.~(\ref{ro1egbis}) are
\begin{eqnarray}
\label{W1eg}  W^{(e),(g)}_{1}(\alpha) &=&2
e^{-2|\alpha|^2} \bigl[ e^{-2|\xi|^2}
\cosh(4|\xi|\mathrm{Im}\,\alpha) \nonumber \\ &&  \pm
\cos(4|\xi|\mathrm{Re}\,\alpha)\bigr]/\bigl(1\pm e^{-2|\xi|^2}\bigr)\,.
\end{eqnarray}
Beyond the two-peaked structure, present in Eq.~(\ref{W1}) for an
unmeasured atom (see also Fig.~\ref{fig1}), the presence of the
sinusoidal interference term implies that the Wigner functions in
Eq.~(\ref{W1eg}) can exhibit strong oscillations with period
$\pi/2|\xi|$. They can even take negative values (see
Fig.~\ref{fig2}), which is a signature of the quantum nature of
the cavity field states of Eq.~(\ref{ro1egbis}). In particular, in
the origin of phase space, $ W^{(e),(g)}_{1}(0) = \pm 2 $.

\begin{figure}[!t]
{\centering
\makebox[80mm][l]{(a)}\\[-3ex] \includegraphics[width=68mm]{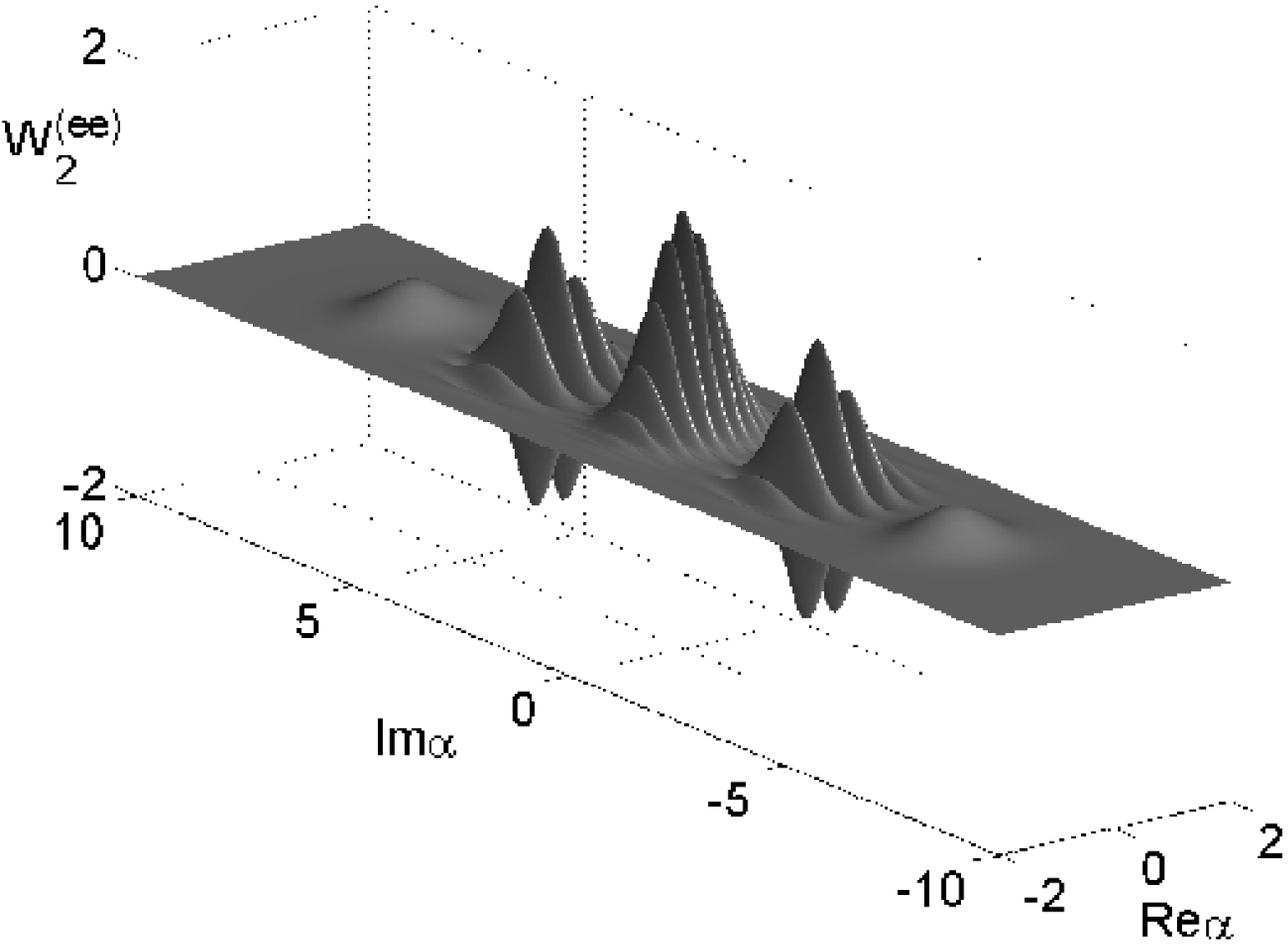}\\
\makebox[80mm][l]{(b)}\\[-3ex] \includegraphics[width=68mm]{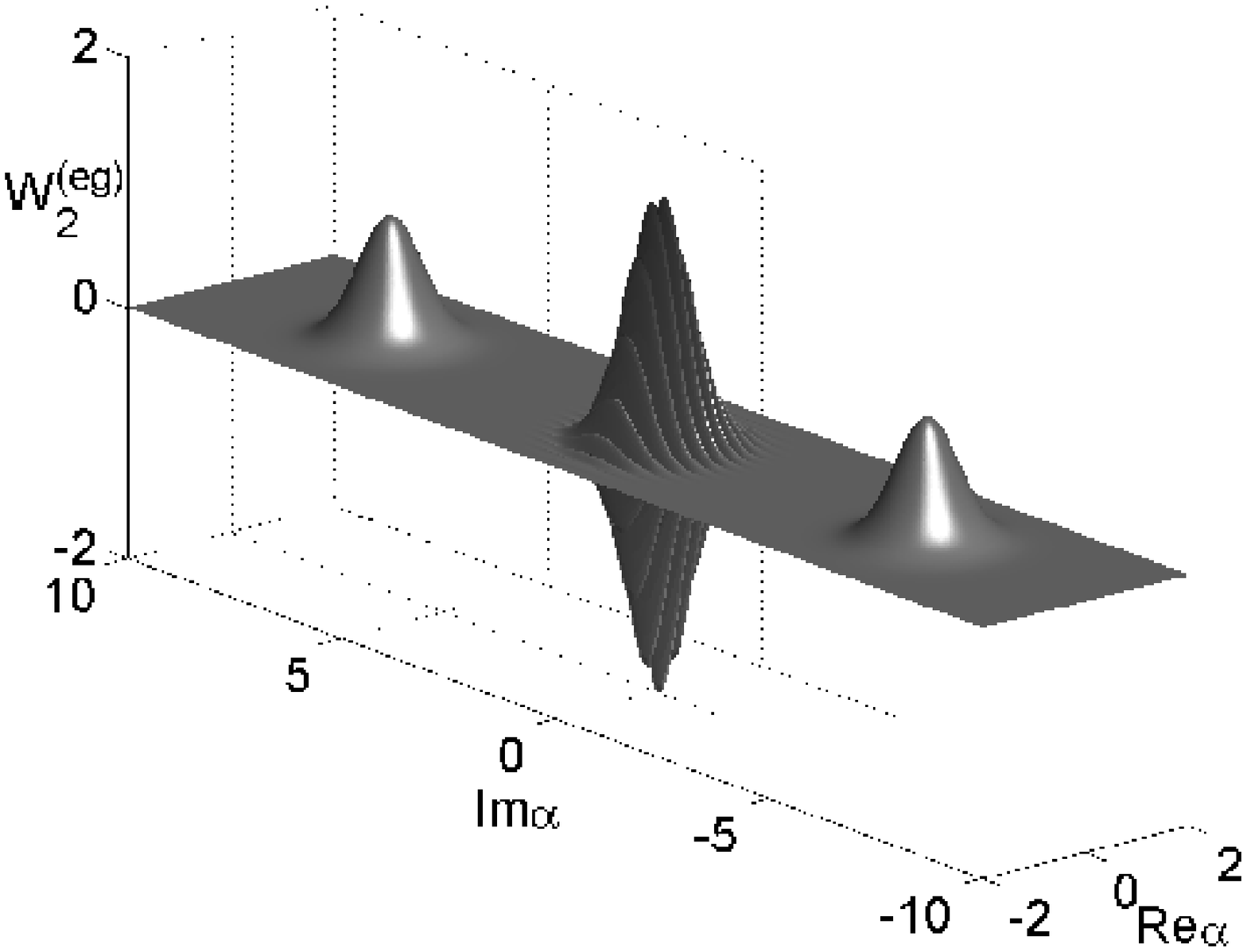}\\
\makebox[80mm][l]{(c)}\\[-3ex] \includegraphics[width=68mm]{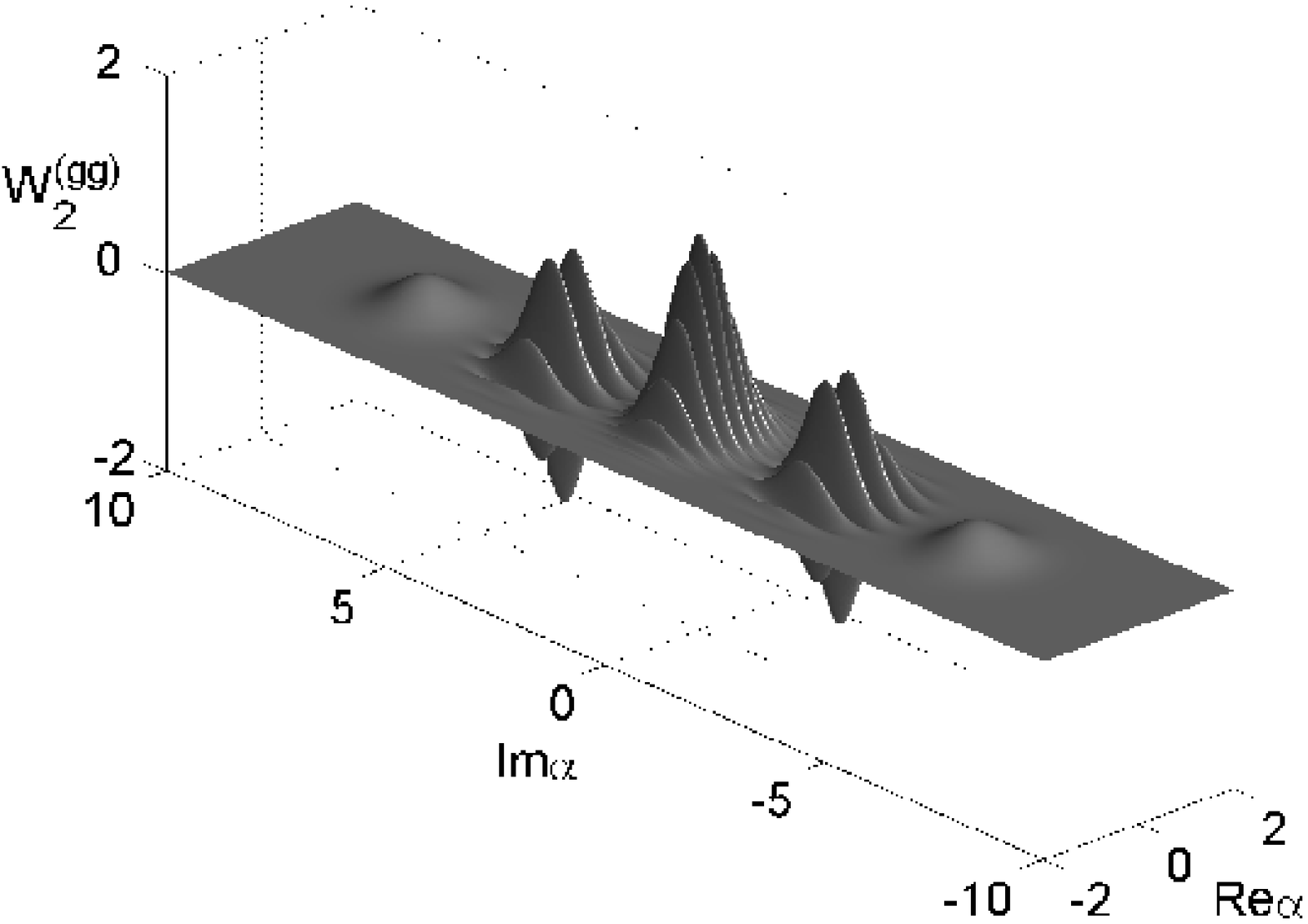}}
\caption{\label{fig3} Wigner distribution of the cavity field for
the case of two crossing atoms detected (a)~both in the excited
state, $W_2^{(ee)} (\alpha)$; (b)~one in the excited state and the
other in the ground state, $W_2^{(eg)} (\alpha)$; (c)~both in the
ground state, $W_2^{(gg)} (\alpha)$; see Eq.~(\ref{W2eg}).}
\end{figure}

If a second atom crosses the cavity and is also detected in the
upper or lower state soon after, the cavity field is projected
onto one of the following states,
\begin{eqnarray}
\label{ro2eg} {\rho}^{(ee)}_{F,2}& = &  \bigl\lbrack |2\xi\rangle
\langle 2\xi| + 4 |0\rangle \langle 0| + |-2\xi\rangle \langle
-2\xi| + |2\xi\rangle \langle -2\xi| \nonumber \\ && +
|-2\xi\rangle \langle 2\xi| + 2 ( |2\xi\rangle \langle 0|+
|0\rangle \langle 2\xi|+ |0\rangle \langle -2\xi| \nonumber \\ &&
+ |-2\xi\rangle \langle 0| ) \bigr\rbrack
\Big/ \bigl\lbrack 2( 3 + 4e^{-2|\xi|^2} + e^{-8|\xi|^2} )
\bigr\rbrack\,,\nonumber \\
{\rho}^{(eg)}_{F,2} &=&  {\rho}^{(ge)}_{F,2} = \bigl\lbrack
|2\xi\rangle \langle 2\xi| + |-2\xi\rangle \langle -2\xi|
\nonumber \\ &&  - |2\xi\rangle \langle -2\xi|- |-2\xi\rangle
\langle 2\xi| \bigr\rbrack
\Big/ \bigl\lbrack 2(1 - e^{-8|\xi|^2})\bigr\rbrack\,, \nonumber \\
{\rho}^{(gg)}_{F,2} &=& \bigl\lbrack |2\xi\rangle \langle 2\xi| + 4
|0\rangle \langle 0| + |-2\xi\rangle \langle -2\xi| + |2\xi\rangle
\langle -2\xi|\nonumber \\ &&  + |-2\xi\rangle \langle 2\xi| - 2 (
|2\xi\rangle \langle 0|+ |0\rangle \langle 2\xi|+ |0\rangle
\langle -2\xi| \nonumber \\ && + |-2\xi\rangle \langle 0| ) \bigr\rbrack
\Big/ \bigl\lbrack 2( 3 - 4e^{-2|\xi|^2} + e^{-8|\xi|^2} )\bigr\rbrack\,,
\nonumber \\
\end{eqnarray}
where, for instance, (eg) means that the first atom is detected in
the upper state and the second one in the lower state. The
expressions of Eq.~(\ref{ro2eg}) should be compared with the
statistical mixture of Eq.~(\ref{roFm}) in the case of two
unobserved atoms ($m=2$). The density operators in
Eq.~(\ref{ro2eg}) describe mesoscopic Schr\"{o}dinger-cat-like
states of the cavity field, the corresponding state vectors being
\begin{eqnarray}
\label{psi2eg} {|{\psi}_{F,2}\rangle}^{(ee),(gg)}&\propto& (|2\xi\rangle
\pm
2|0\rangle + |-2\xi\rangle)\,, \nonumber \\
{|{\psi}_{F,2}\rangle}^{(eg)} = {|{\psi}_{F,2}\rangle}^{(ge)}&\propto&
(|2\xi\rangle - |-2\xi\rangle)\,.
\end{eqnarray}
The Wigner functions which represent the above states in phase
space, whose behavior is depicted in Fig.~\ref{fig3}, can be
written as
\begin{eqnarray}
\label{W2eg} W^{(ee)}_{2}(\alpha) &=& 2 e^{-2|\alpha|^2}
\bigl\lbrack 2 + e^{-8|\xi|^2} \cosh(8|\xi|\mathrm{Im}\,\alpha)
  \nonumber \\ &&
+ 4e^{-2|\xi|^2} \cosh(4|\xi|\mathrm{Im}\,\alpha)
\cos(4|\xi|\mathrm{Re}\,\alpha) \nonumber \\&&
+ \cos(8|\xi|\mathrm{Re}\,\alpha)\bigr\rbrack \Big /
 \bigr( 3 + 4e^{-2|\xi|^2} + e^{-8|\xi|^2} \bigr) \,,
\nonumber \\
W^{(eg)}_{2}(\alpha)&=& W^{(ge)}_{2}(\alpha) = 2 e^{-2|\alpha|^2}
\bigl\lbrack - \cos(8|\xi|\mathrm{Re}\,\alpha) \nonumber \\ &&  +
e^{-8|\xi|^2} \cosh(8|\xi|\mathrm{Im}\alpha) \bigr\rbrack\Big/
\bigl(1 - e^{-8|\xi|^2} \bigr)
\,,\nonumber \\
 W^{(gg)}_{2}(\alpha) &=&
2 e^{-2|\alpha|^2}
\bigl\lbrack 2   + e^{-8|\xi|^2} \cosh(8|\xi|\mathrm{Im}\,\alpha) \nonumber \\
&& - 4e^{-2|\xi|^2} \cosh(4|\xi|\mathrm{Im}\alpha)
\cos(4|\xi|\mathrm{Re}\,\alpha)\nonumber\\
&&  + \cos(8|\xi|\mathrm{Re}\,\alpha)
\bigr\rbrack \Big/
 \bigl( 3 - 4e^{-2|\xi|^2} + e^{-8|\xi|^2}
\bigr)\,.\nonumber\\
\end{eqnarray}

We consider now the atomic statistics at the exit of the cavity
independent of the cavity field state. If an atom is injected in
the cavity, so that the initial state is $\rho_{F}\otimes
|e\rangle\langle e|$, the atomic state after the interaction will
be
\begin{eqnarray}
\label{roA} \rho_{A} &=& \mathrm{Tr}_{F} \bigl\{ U(\xi) \rho_{F}\otimes
|e\rangle\langle e| U(-\xi) \bigr\} \nonumber \\
 &=& \frac{1}{2}\Bigl\lbrack 1_{A} + \mathrm{Re}\,\chi(2\xi)
\bigl(|e\rangle\langle e|-|g\rangle\langle g|\bigr) \nonumber \\
&&+ i\, \mathrm{Im}\,\chi(2\xi)\bigl(|g\rangle\langle e|-
 |e\rangle\langle g|\bigr) \Bigr\rbrack\,.
\end{eqnarray}
Here, $\mathrm{Tr}_{F}$ is the partial trace over the cavity mode
degrees of freedom. In the derivation of Eq.~(\ref{roA}) we used
also Eqs.~(\ref{U}) and (\ref{ro0}) as well as the property
$\chi(\beta) = \chi(-\beta)^*$. Finally, we can calculate the
probability $p_{e,g}$ for atomic detection in the upper or lower
state as
\begin{eqnarray}
\label{peg} p_{e,g} = \langle e,g| \rho_{A} |e,g\rangle =
\frac{1}{2} \bigl[ 1 \pm \,\mathrm{Re}\,\chi(2\xi) \bigr]\,.
\end{eqnarray}

\section{Master equation and analytical results}\label{sec3}

\subsection{SDM master equation and analytical solutions}\label{sec3a}

We consider now the case of a poissonian beam of two-level Rydberg
atoms, with pumping rate $r$, interacting with a single mode of a
microwave high-Q cavity. While the atoms are inside the cavity,
during an interaction time $\tau$, they are strongly driven by an
additional classical field, the whole system being in resonance.
We study a regime where at most one atom is present inside the
cavity, that is, $\tau \ll r^{-1}$, and where the decay of atomic
Rydberg levels is negligible. Between two successive atoms, the
cavity field decays due to its interaction with a thermal bath.
Hence, even though the present strongly-driven micromaser (SDM)
looks quite similar to the conventional micromaser, there is a crucial
difference in the pumping dynamics. As we discussed in the
previous section, the JC interaction and its Rabi oscillations do
not rule any more the unitary atom-field interaction.
Their role is taken over by field displacements conditioned on
 the atomic internal states.

The dynamics of the SDM field is ruled by the interplay between
the amplification process, due to the interaction with the driven
atoms, and the dissipation process, occurring when no atom crosses
the cavity. In a coarse-grained description the gain rate of the
cavity mode follows from Eq.(\ref{roF1}),
\begin{equation}
\label{gain} \frac{\partial\rho_{F}}{\partial t}\Bigg|_\mathrm{gain} =
\frac{r}{2}\bigl[ D(\xi)\rho_{F} D(-\xi) + D(-\xi)\rho_{F} D(\xi) -
2\rho_{F} \bigr]\, .
\end{equation}
The loss rate, due to the interaction of the cavity field with a
thermal bath~\cite{SZ}, is given by
\begin{eqnarray}
\label{loss}\frac{\partial\rho_{F}}{\partial t}\Bigg|_\mathrm{loss} & = &
-\frac{\gamma (\bar{n}+1)}{2} \bigl[ a^\dagger a\rho_{F} - 2a\rho_{F}
a^\dagger + \rho_{F} a^\dagger a\bigr]
\nonumber \\
&& - \frac{\gamma \bar{n}}{2} \bigl[ aa^\dagger\rho_{F} -
2a^\dagger\rho_{F} a + \rho_{F} aa^\dagger\bigr]  \equiv
\mathcal{L}\rho_{F}\,, \nonumber \\
\end{eqnarray}
where $\gamma$ is the cavity photon decay rate and $\bar{n}$ the
mean thermal photon number. By combining Eqs.~(\ref{gain}) and
(\ref{loss}) we obtain the SDM master equation
\begin{equation}
\label{ME} \frac{\partial\rho_{F}}{\partial t} =
\frac{r}{2} \lbrack D(\xi)\rho_{F} D(-\xi) + D(-\xi)\rho_{F}
D(\xi) - 2\rho_{F} \rbrack + \mathcal{L}\rho_{F}\,.
\end{equation}
At variance with the conventional micromaser, whose master
equation has only an analytical solution in
steady-state~\cite{Meystre}, the SDM master equation (\ref{ME})
can be fully solved for any time. The key step is a
transformation of the equation of motion (\ref{ME}) for the density
operator into the
corresponding partial differential equation for its symmetrically
ordered characteristic function $\chi(\beta)$, already introduced
in the previous section. We recall that $\chi(\beta)$ is related
to the field density operator by~\cite{CG}
\begin{equation}
\label{reverse} \rho_{F} =
\pi^{-1}\int\chi(\beta)D^{-1}(\beta)\,d^2\beta\,
\end{equation}
and that the Wigner function, $W(\alpha)$, is the 2D Fourier
transform of the characteristic function $\chi(\beta)$~\cite{CG}.

Using known operator techniques~\cite{SZ}, we thus map
Eq.~(\ref{ME}) into the following partial differential equation
for $\chi(\beta ,\beta^* ,t)$,
\begin{eqnarray}
\label{masterchi} \frac{\partial}{\partial t}
\chi(\beta ,\beta^* ,t) &=& \Bigl[\frac{r}{2}\bigl(e^{\xi^*
\beta - \xi \beta^*} + e^{-\xi^*
\beta + \xi \beta^*} - 2 \bigr) \nonumber \\
&&  -\frac{\gamma}{2}(2\bar{n}+1)|\beta|^2\Bigr]
\chi(\beta ,\beta^* ,t) \nonumber \\
&& \mbox{} - \frac{\gamma}{2}\Bigl(\beta \frac{\partial}{\partial
\beta} + \beta^* \frac{\partial}{\partial\beta^*}\Bigr)\chi(\beta ,
\beta^* ,t)\,.\nonumber\\
\end{eqnarray}
We transform Eq.~(\ref{masterchi}) to cartesian coordinates by
setting $\xi^*\beta \equiv|\xi|(x+iy)$, which gives
\begin{equation}
\label{cartesian}\biggl(\frac{\partial}{\partial t}
+\frac{\gamma}{2}\Bigl[x\frac{\partial}{\partial x} +
y\frac{\partial}{\partial y} +xG^{\prime}(x) + yF^{\prime}(y)
\Bigr]\biggr)\chi(x,y,t)=0\, ,
\end{equation}
where
\begin{eqnarray}
\label{Nex}G(x) & = & \frac{2\bar{n}_b +1}{2} x^2 \,, \nonumber \\
F(y) & = & \int\limits^y_0 \frac{4N_{ex}\sin^2(|\xi|z)+(2\bar{n}_b
+1)z^2}{z} dz\, ,
\end{eqnarray}
and $N_{ex} = r/{\gamma}$. The solution of Eq.~(\ref{cartesian})
reads
\begin{equation}
\label{center}\chi(x,y,t) = \frac{\chi^\mathrm{(ss)}(x,y)\chi_{0}
(xe^{-\frac{\gamma t}{2}},ye^{-\frac{\gamma t}{2}})}
{\chi^\mathrm{(ss)}(xe^{-\frac{\gamma t}{2}},ye^{-\frac{\gamma t}{2}})} .
\end{equation}
Here,
\begin{equation}
\label{sssolution}\chi^\mathrm{(ss)}(x,y)=e^{-G(x)-F(y)}
\end{equation}
is the steady-state solution of Eq.~(\ref{cartesian}) that is reached for
$t\rightarrow \infty$, and $\chi_{0}=\mathrm{Tr}_{F}[\rho_{F}(0)D]$ is the
characteristic function corresponding to the initial field state
$\rho_{F}(0)$. By substituting the expressions of Eq.~(\ref{Nex})
into Eq.~(\ref{sssolution}), we arrive at the steady-state
characteristic function
\begin{eqnarray}
\label{stst}\chi^\mathrm{(ss)}(x,y)&=&  \exp \Bigl( -
(\bar{n}+{\textstyle\frac{1}{2}}) (x^2 + y^2) \nonumber \\ && +
4N_{ex}\bigl[-\gamma_{e}-\ln(2|\xi|y)+ \mathrm{Ci}
(2|\xi|y)\bigr]\Bigr)\, ,
\nonumber\\
\end{eqnarray}
where $\gamma_{e}$ is Euler's constant and $\mathrm{Ci}(2|\xi|y)$
is the Cosine Integral defined as
\begin{equation}
\mathrm{Ci}(2|\xi|y)=\gamma_{e} + \ln(2|\xi|y) + \int\limits^{2|\xi|y}_0
\frac{\cos(z) -1}{z}dz\,.
\end{equation}
Note that starting from a real initial function, the
time-dependent solution in Eq.~(\ref{center}) will always produce
a real characteristic function. This follows from the invariance
of the time evolution equation in Eq.~(\ref{masterchi}) under the
transformation $\beta \rightarrow -\beta$, and from the property
$\chi(\beta) = \chi(-\beta)^*$. Accordingly, the solution of
Eq.~(\ref{center}) depends only on the modulus of the (imaginary)
parameter $\xi$, which is in agreement with the invariance of the
master equation (\ref{ME}) under the transformation
$\xi\rightarrow -\xi$.

\subsection{SDM field statistics}\label{sec3b}

It is clear that from the general solution of the characteristic
function, Eq.~(\ref{center}), we can calculate the field density
operator $\rho_{F}(t)$, following Eq.~(\ref{reverse}), and have
access to the field statistics. However, this information can be
directly extracted from the characteristic function itself, since
the expectation value of any symmetrically ordered product of the
operators $a$ and $a^{\dagger}$ is given by~\cite{CG}
\begin{equation}
\label{moments}\langle(a^{\dagger})^{m}a^{n}\rangle_\mathrm{sym} =
\frac{\partial^{m+n}\chi(\beta,\beta^{*})}{\partial\beta^{m}
\partial(-\beta^{*})^{n}}\bigg|_{\beta=\beta^{*}=0}.
\end{equation}
For example, we have
\begin{eqnarray}
\label{meanvalues}
\langle a(t)\rangle & = & \langle a(0)\rangle
e^{-\frac{\gamma}{2}t}\,, \nonumber \\
\langle a^{\dagger}(t)a(t)\rangle & = & \langle a^{\dagger}
(0)a(0)\rangle e^{-\gamma t} + (N_{ex}|\xi|^{2}+
\bar{n})(1-e^{-\gamma t})\,, \nonumber \\
\langle a^{2}(t)\rangle & = & \langle a^{2}(0)\rangle
e^{-\gamma t}+ N_{ex} \xi^{2} (1-e^{-\gamma t})\,.
\end{eqnarray}
From the expressions in Eq.~(\ref{meanvalues}) and their complex
conjugates, we derive the steady-state expectation values of the
field amplitude, photon number, and quadrature variances $(\Delta
x_{1,2})^{2}$, with $x_{1} = (1/2)(a^{\dagger} + a)$ and $x_{2} =
(i/2)(a^{\dagger} - a)$,
\begin{eqnarray}
\label{ssmeanvalues}
\langle a\rangle^\mathrm{(ss)} & = & 0 \,,  \nonumber \\
\langle a^{\dagger}a\rangle^\mathrm{(ss)} & = & N_{ex}|\xi|^{2} + \bar{n}
\,,  \nonumber \\
({\Delta x_{1}}^\mathrm{(ss)})^{2}
& = & {\textstyle\frac{1}{4}} (1 + 2\bar{n})\,, \nonumber \\
({\Delta x_{2}}^\mathrm{(ss)})^{2}
& = & {\textstyle\frac{1}{4}} (1 + 2\bar{n} + 4N_{ex}|\xi|^{2})\,.
\end{eqnarray}
We see that at steady-state the expectation value of the SDM field
is zero, whereas the mean photon number is a quadratic function of
the modulus of the single atom displacement parameter $\xi$. Also,
for vanishing cavity temperature $(\bar{n}\rightarrow 0)$, the
variance of the quadrature operator $x_{1}$ remains at the minimum
value $1/4$, whereas the variance of the orthogonal quadrature
$x_{2}$, the one that is being driven by the system, is broadened
by a factor equal to $\langle a^{\dagger}a\rangle^\mathrm{(ss)}$.

Another important quantity describing the photon statistics is the
Fano-Mandel parameter $Q = [(\Delta N)^{2} - \langle
N\rangle]/\langle N\rangle$, where $N \equiv \langle
a^{\dagger}a\rangle$. For the SDM at steady-state we obtain
\begin{equation}
\label{Q} Q = 1 + N_{ex}|\xi|^{2}+\bar{n} +
\frac{N_{ex}|\xi|^{4}(N_{ex}+\frac{1}{2})}
{N_{ex}|\xi|^{2}+\bar{n}},
\end{equation}
which describes a persistent superpoissonian behavior $(Q
> 1)$. This result is at variance with the conventional micromaser
steady-state~\cite{Meystre}, which alternates between
superpoissonian and subpoissonian statistics.

\subsection{Atomic correlations}\label{sec3c}

Just as it is the case for the micromaser, in the SDM we are not able to
measure the cavity field directly. Therefore, we use the atoms not
only for pumping the cavity mode but also as a source of
information about the SDM field. The theory of the detector clicks
statistics as well as the connection between the statistics of the
detected atoms and the cavity field was developed for the
conventional micromaser in Ref.~\cite{Englert2}. According to this
approach, the detection of the exiting atom, initially excited, in
the ground state is described by an operator ${\mathcal A}$ whose
action on the field density operator $\rho_{F}$ is
\begin{eqnarray}
{\mathcal A} \rho_{F} = \frac{1}{4}\bigl[ D(-\xi)\rho_{F} D(\xi) +
D(\xi)\rho_{F} D(-\xi) \nonumber \\ -  D(-\xi)\rho_{F} D(-\xi) -
D(\xi)\rho_{F} D(\xi) \bigr]\,.
\end{eqnarray}
Likewise, the click operator $\mathcal{B}$ for detection in the
excited state acts as follows
\begin{eqnarray}
{\mathcal B}\rho_{F} = \frac{1}{4}\bigl[ D(-\xi)\rho_{F} D(\xi) +
D(\xi)\rho_{F} D(-\xi) \nonumber \\ + D(-\xi)\rho_{F} D(-\xi) +
D(\xi)\rho_{F} D(\xi) \bigr]\,.
\end{eqnarray}
Note that the above expressions have the same structure as the
field state operators of Eqs.~(\ref{ro1eg}). The probability to
detect the atom in the ground (excited) state $p_{g}$ ($ p_{e}$)
can be calculated as $\mathrm{Tr}_{F}[{\mathcal A}\rho_{F}]$ ($
\mathrm{Tr}_{F}[{\mathcal B}\rho_{F}]$), giving the results of
Eq.~(\ref{peg}), which can be further simplified after the
derivation of a real expression for the characteristic function
$\chi$, so that
\begin{equation}
p_{e,g}  = \frac{1}{2}[ 1 \pm \chi(2\xi) ]\,.
\end{equation}
The introduction of the above click operators enables us to
describe the atomic correlation functions, or conditional
probabilities for two consecutive detector clicks. For instance,
the correlation function for a detection of the first atom in the
excited state after a detection of the second atom in the ground
state separated by a time interval $t$ is given by
\begin{equation}
\label{ge}G_{ge}(t) = \frac{\mathrm{Tr}_{F}[{\mathcal B}e^{{\mathcal
L}_{0}t}{\mathcal A}\rho^\mathrm{(ss)}]} {\mathrm{Tr}_{F}[{\mathcal
B}\rho^\mathrm{(ss)}]\mathrm{Tr}_{F}[{\mathcal A}\rho^\mathrm{(ss)}]}
= \frac{1 +
\tilde{\chi}(2\xi,t)}{1 + \chi^\mathrm{(ss)}(2\xi)}\,,
\end{equation}
where ${\mathcal L}_{0} = {\mathcal L} + r({\mathcal A}+{\mathcal
B}-1)$, $\chi^\mathrm{(ss)}$ is the steady-state characteristic
function(\ref{stst}), and
\begin{eqnarray}
\tilde{\chi}(2\xi,t) &=&
\frac{2\chi^\mathrm{(ss)}(2\xi e^{-\frac{A}{2}t})
- 1 - \chi^\mathrm{(ss)}(4\xi
e^{-\frac{A}{2}t})}{2 - 2\chi^\mathrm{(ss)}(2\xi e^{-\frac{A}{2}t})}
\nonumber \\ && \times \exp \Bigl(
-(\bar{n}+{\textstyle\frac{1}{2}}) |2\xi|^{2}(1-e^{-\frac{A}{2}t})
\Bigr)\,.
\end{eqnarray}

\begin{figure}
\makebox[80mm][l]{$t=0$}\\
\centerline{\includegraphics[width=65mm]{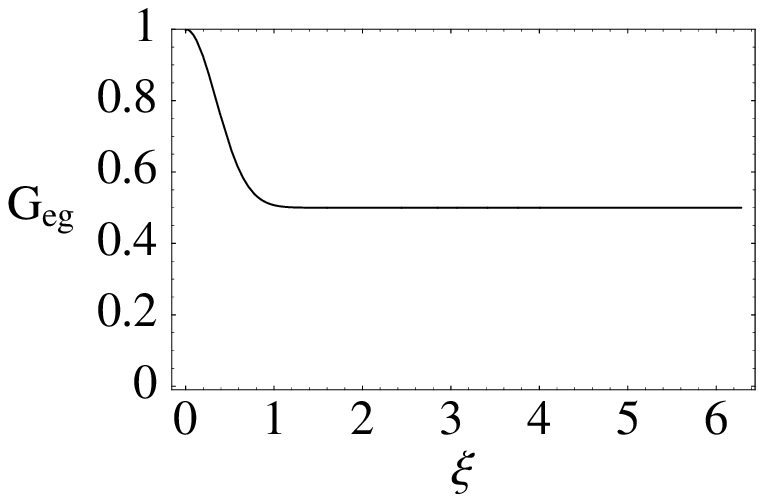}}
\makebox[80mm][l]{$t=0.01$}\\
\centerline{\includegraphics[width=65mm]{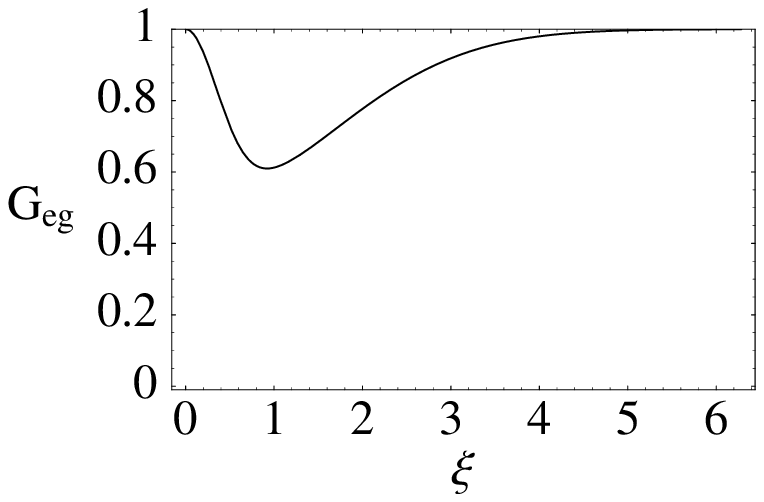}}
\caption{\label{fig4} Atomic correlations $G_{eg}$ with parameters
$N_{ex} = 50$, $\bar{n}=0.03$. The figures are plotted for time
intervals $t=0$ (top) and $t=0.01$ (bottom) between detection
clicks.}
\end{figure}

The two-click correlation function in Eq.~(\ref{ge}) is the ratio
between the conditional probability to have a second $e$-click at
time $t$ after a first $g$-click occurred at time $0$, and the
probability for an $e$-click when the cavity field is in the
steady-state. All other correlation functions for click pairs
$G_{ee}$, $G_{gg}$ and $G_{eg}$ are given by the following
expressions
\begin{eqnarray}
\label{other}  G_{gg}(t) &=&
\frac{\mathrm{Tr}_{F}[{\mathcal A}e^{{\mathcal L}_{0}t}
{\mathcal A}\rho^\mathrm{(ss)}]}
{\bigl(\mathrm{Tr}_{F}[{\mathcal A}\rho^\mathrm{(ss)}]\bigr)^{2}} =
\frac{1 - \tilde{\chi}(2\xi,t)}{1 - \chi^\mathrm{(ss)}(2\xi)} \,,
\nonumber \\
G_{eg}(t) &=&
\frac{\mathrm{Tr}_{F}[{\mathcal A}e^{{\mathcal L}_{0}t} {\mathcal
B}\rho^\mathrm{(ss)}]}{\mathrm{Tr}_{F}[{\mathcal A}
\rho^\mathrm{(ss)}]\mathrm{Tr}_{F} [{\mathcal
B}\rho^\mathrm{(ss)}]} = \frac{1 - \bar{\chi}(2\xi,t)}{1 -
\chi^\mathrm{(ss)}(2\xi)}\,,\nonumber  \\
G_{ee}(t) &=&
\frac{\mathrm{Tr}_{F}[{\mathcal B}e^{{\mathcal L}_{0}t}{\mathcal B}
\rho^\mathrm{(ss)}]}
{\bigl(\mathrm{Tr}_{F}[{\mathcal B}\rho^\mathrm{(ss)}]\bigr)^{2}}
= \frac{1 +\bar{\chi}(2\xi,t)}{1 + \chi^\mathrm{(ss)}(2\xi)}\,,
\end{eqnarray}
where
\begin{eqnarray}
\bar{\chi}(2\xi,t) &=& \frac{ 2\chi^\mathrm{(ss)}(2\xi
e^{-\frac{A}{2}t}) + 1 + \chi^\mathrm{(ss)}(4\xi e^{-\frac{A}{2}t}) }{ 2 +
2\chi^\mathrm{(ss)}(2\xi e^{-\frac{A}{2}t}) } \nonumber \\ &&
\times \exp \Bigl( -(\bar{n}+{\textstyle\frac{1}{2}}) |2\xi|^2
(1-e^{-\frac{A}{2}t}) \Bigr)\,.
\end{eqnarray}

However, like in the case of the conventional micromaser, the
two-click correlation functions of Eqs.~(\ref{ge}) and
(\ref{other}) for the SDM reflect the statistics of all possible
consecutive two-click events separated by a time $t$. In order to
judge about the correlations between two truly successive atoms,
one should take the limit $t\rightarrow 0$ in Eqs.~(\ref{ge}) and
(\ref{other}). In this limit, two consecutive atoms interact with
the cavity field and there is no time for decoherence to take
place in between. Therefore, one expects stronger atomic
correlations as a consequence of the discussion done in
Sec.~\ref{sec2b}

In Fig.~\ref{fig4}, we show the correlation function $G_{eg}$ as
a function of the displacement $\xi$. We see that $G_{eg}$
exhibits stronger correlations, when compared with the
conventional micromaser, despite the complete classicality of the
SDM steady-state. This is a counter intuitive result coming from
the belief that stronger correlations should appear only when
non-classical steady-states are involved. The stronger atomic
correlations present in the SDM originate in the different nature
of the unitary process. In the SDM, we rely on an interaction (see
Sec.~\ref{sec2}) that naturally produces mesoscopic entangled
atom-field states (``Schr\"odinger cat states''), while the
conventional micromaser uses the Jaynes-Cummings interaction,
producing Rabi oscillations inside well defined atom-field
subspaces, and essentially exchanging a single photon per cycle.

\section{Numerical results}\label{sec4}

In order to describe the SDM dynamics we have at our disposal the
analytical expression of the symmetrically ordered characteristic
function $\chi(\beta ,\beta^* ,t)$, Eq.(\ref{center}). We can as
well describe the time evolution of the Wigner function
$W(\alpha,\alpha^*,t)$, that is the Fourier transform of
$\chi(\beta,\beta^*,t)$, which gives a picture in the phase space
associated to the cavity mode. We can further consider the time
behavior of the density matrix $(\rho_{F})_{m,n}(t) = \langle
m|\rho_{F}(t)|n\rangle$, derived from the master equation in
Eq.~(\ref{ME}) when the field density operator is represented in
the Fock basis. In this case, we can use quantum jumps techniques
~\cite{QT}, already applied succesfully in several problems in the
domain of CQED~\cite{CL,CL2}.

\begin{figure}
{\centering
\makebox[80mm][l]{(a)}\\[-3ex] \includegraphics[width=68mm]{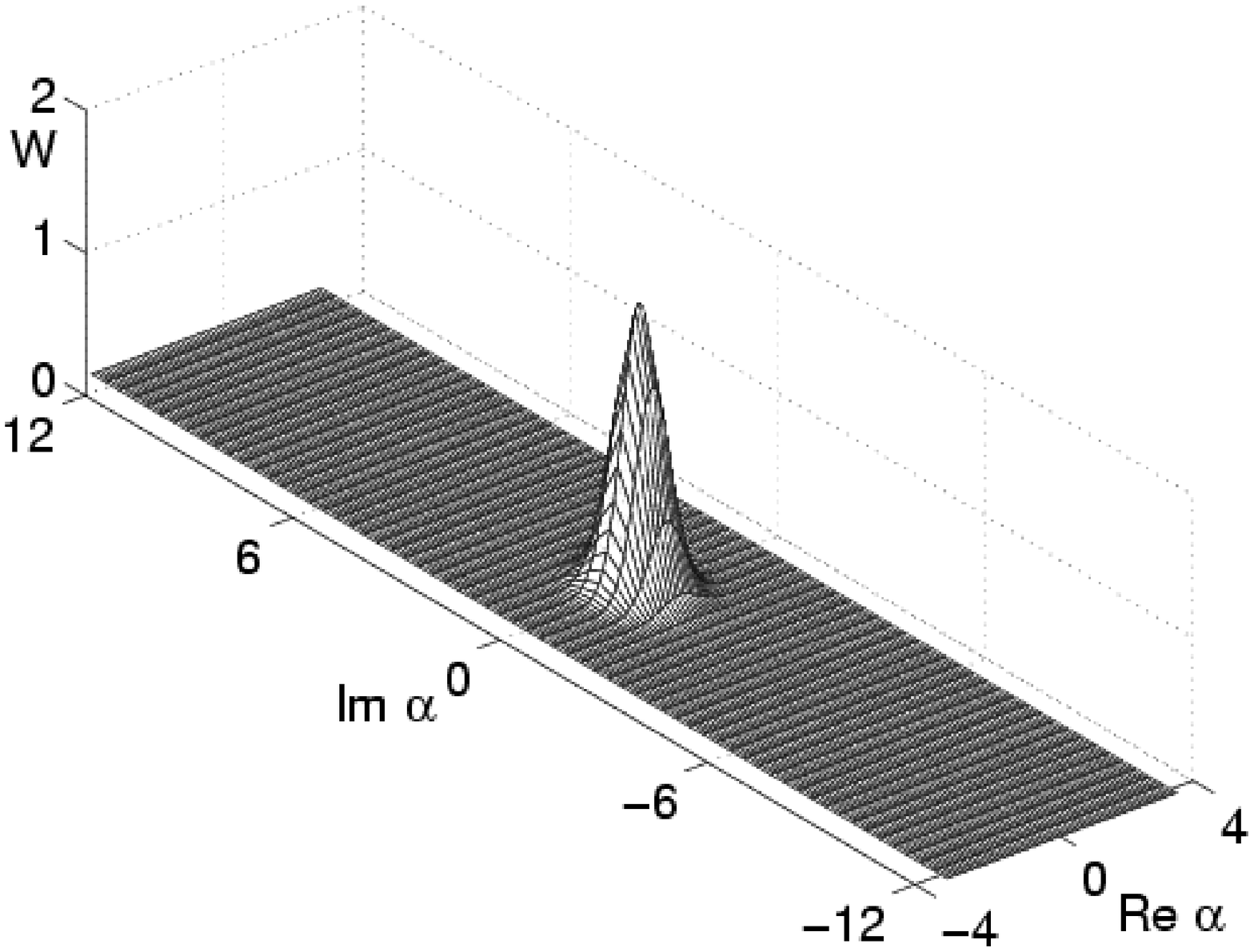}\\
\makebox[80mm][l]{(b)}\\[-3ex] \includegraphics[width=68mm]{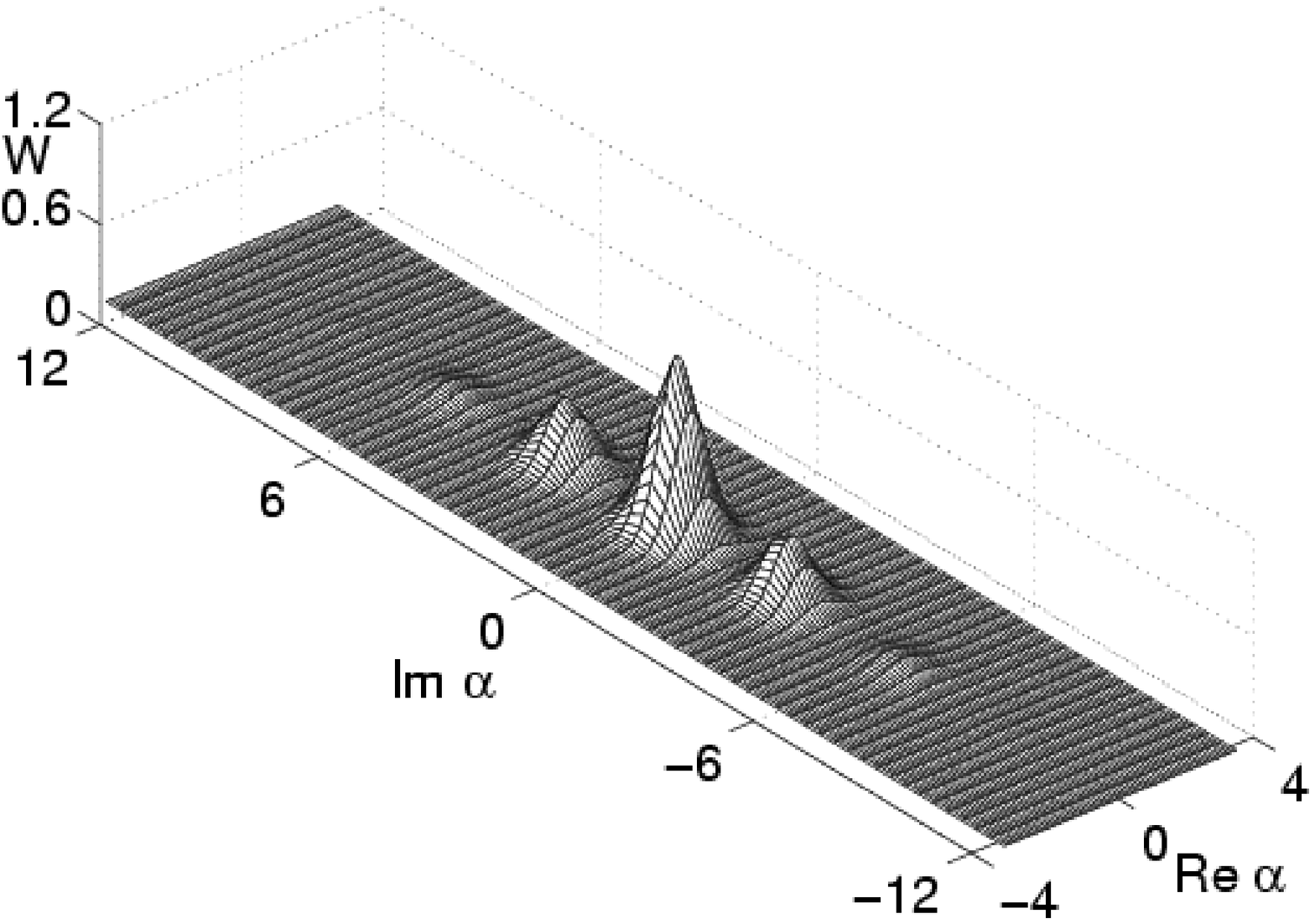}\\
\makebox[80mm][l]{(c)}\\[-3ex] \includegraphics[width=68mm]{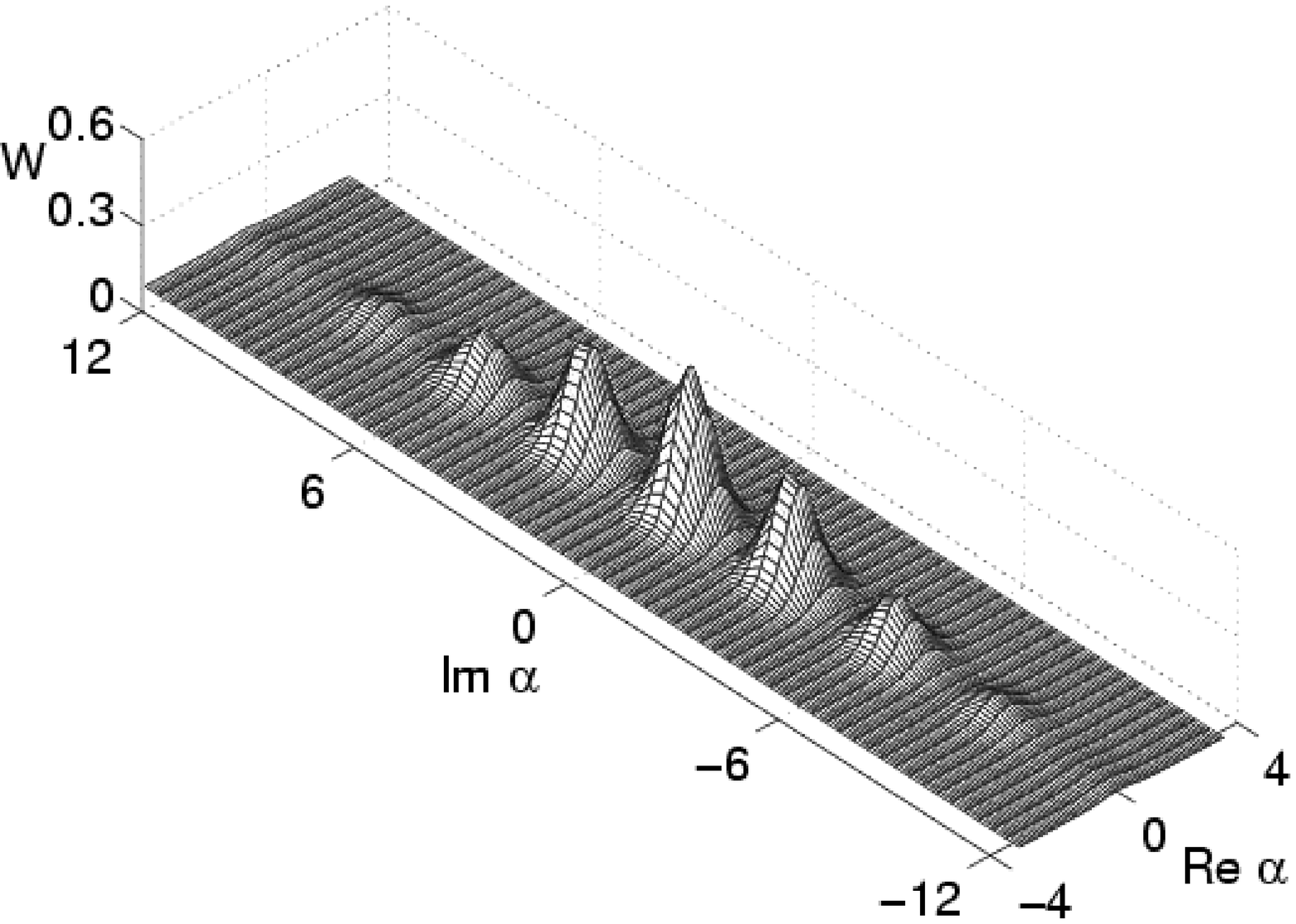}\\
\makebox[80mm][l]{(d)}\\[-3ex] \includegraphics[width=68mm]{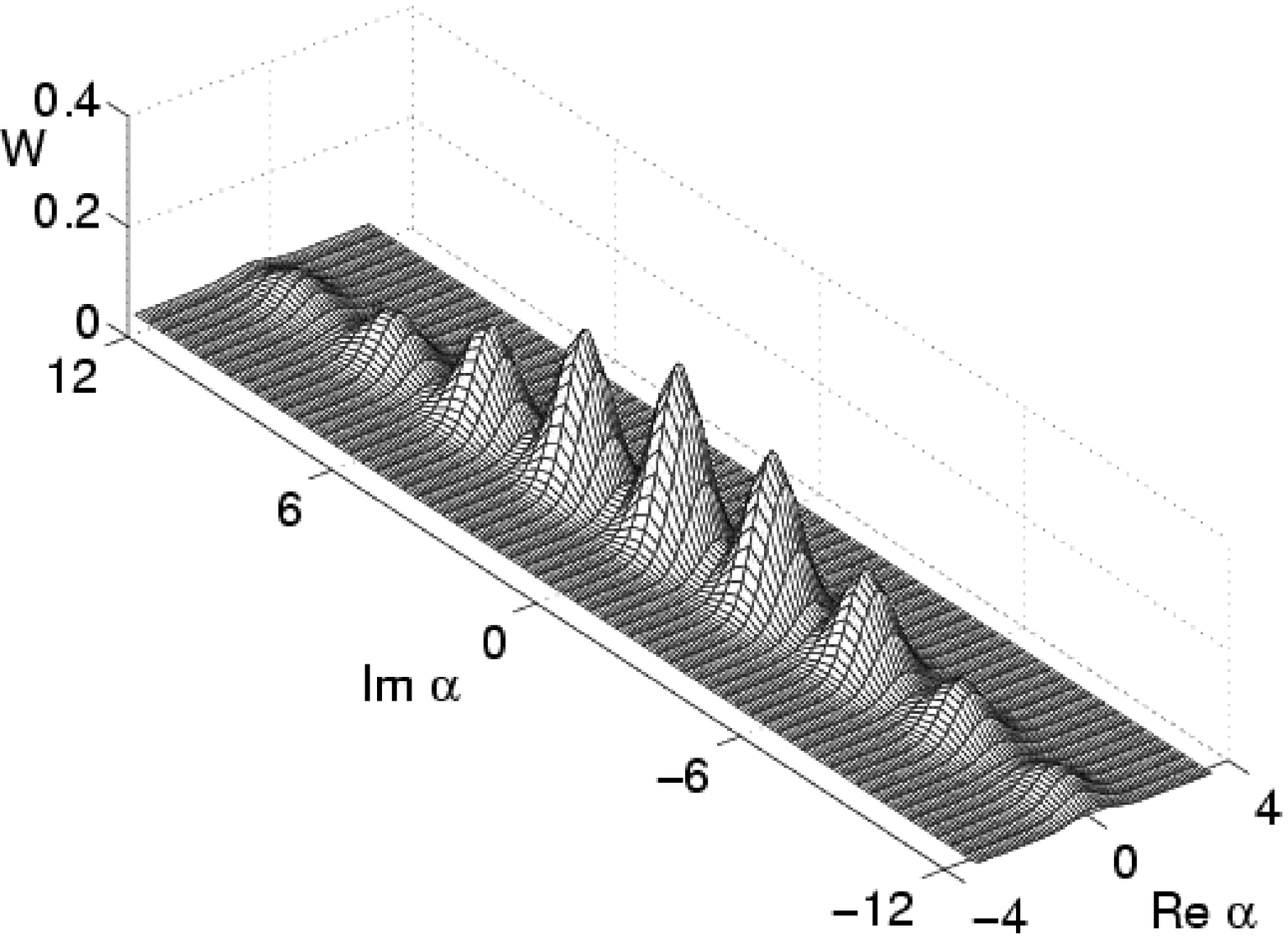}}
\caption{\label{fig5}Generation of  multi-peaked distributions in
the transient cavity field dynamics: $W(\alpha, t)$ distributions
at $t = 0$ (top) and $\gamma t = 0.1, 0.3, 0.5$ (proceeding downward),
obtained from the master equation (\ref{ME}).}
\end{figure}

All these intertwined tools allow a consistent description of SDM
dynamics which complements the analytical results and unveils
additional features. As an example, in Fig.~\ref{fig5} we show the
transient behavior of the Wigner distribution $W(\alpha)$ of the
SDM field. Starting from the Gaussian function of the vacuum state
(Fig.~\ref{fig5}a), after a time interval $\gamma t = 0.1$
(Fig.~\ref{fig5}b) we see the presence of two additional peaks,
symmetrically placed along the imaginary axis. Here, we have
chosen $|\xi| = \pi$, a large enough value to see a peaked
structure. At later times in the transition, Fig.~\ref{fig5}(c,d),
we see the onset of other peak pairs symmetrically placed on the
imaginary axis, while the distribution lowers and broadens along
that axis. The underlying physical mechanism for the generation of
this dynamics was described in Sec.~\ref{sec2}.

\begin{figure}
\centerline{\includegraphics[width=78mm]{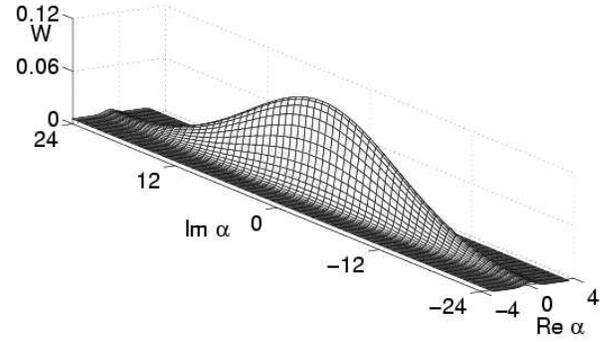}}
\caption{\label{fig6} Steady-state cavity field after the
transient evolution of Fig.~\ref{fig5}: $W(\alpha)$ function at $\gamma t =
20$.}
\end{figure}

\begin{figure}
\centerline{\includegraphics[width=75mm]{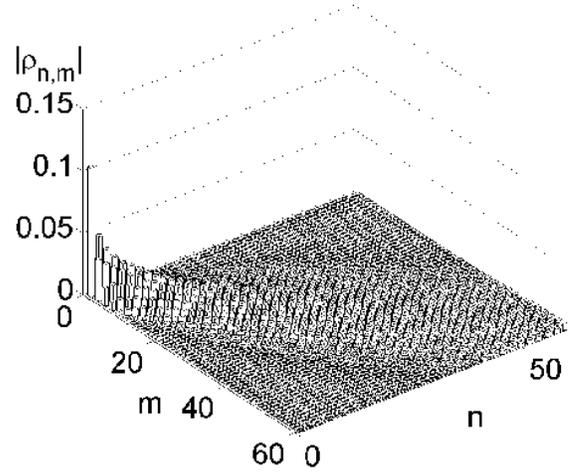}}
\caption{\label{fig7} Density matrix of the steady-state cavity
field.}
\end{figure}

At later times decoherence destroys such structures, and when the
steady-state is reached (Fig.~\ref{fig6}) the distribution looks
like the envelope of the multiply-peaked structure along the
imaginary axis, while it preserves its initial minimum width along
the real axis. Figure~\ref{fig7} shows the density matrix of the
steady-state SDM field. The diagonal elements provide the photon
statistics, and their wide distribution confirms the predicted
super-poissonian behavior. Furthermore, we note the presence of
off-diagonal density matrix elements or coherences, which rule the
phase and spectral properties of the field.

\begin{figure}
\centering{\includegraphics[width=60mm]{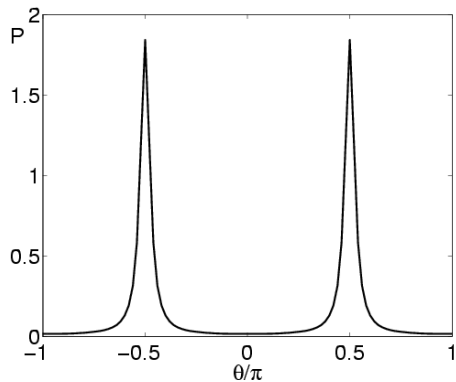}}
\caption{\label{fig8} Pegg-Barnett phase distribution of the
steady-state cavity field.}
\end{figure}

 Initially they are not excited
in the cavity field, hence they are induced by the interaction
with the strongly driven atoms. In order to investigate this
effect, in Fig.~\ref{fig8}, we show the steady-state Pegg-Barnett
\cite{PB} phase distribution
\begin{equation}
\label{phase} P(\theta) = (2\pi)^{-1} \sum_{m,n=0}^{\infty}
{\rho_{F}}^\mathrm{(ss)}_{n,m} e^{i(m-n)\theta}\,.
\end{equation}

The SDM phase distribution shows a particular feature, i.e., a
narrow two-peaked structure centered on the values $\theta = \pm
\pi/2$, which is already present in the transient. This feature is
explained by the SDM gain dynamics, Eq.~(\ref{gain}), which shows
that the resonant interaction of the cavity field with strongly
driven atoms implies an equal displacement of the field by the
imaginary quantities $\pm \xi$. In turn, this effect is quite
consistent with the narrow elongated form of the steady-state
distribution along the imaginary axis in phase space, as well as
with the vanishing of the steady-state field expectation value.

\section{Conclusions}\label{sec5}

We have investigated the dynamics of a strongly-driven micromaser,
based on the resonant interaction of one mode of a high-Q cavity
with a poissonian low density beam of two-level atoms strongly
driven by a resonant classical field. We have shown that this
system provides a nontrivial example of an open quantum  system that
can be described analytically, exhibiting classical and
nonclassical regimes and properties. We have presented the
time-dependent solution of the master equation and the expressions
of the main statistical quantities for both the cavity field and
the detected atoms. In particular, we have found that the
steady-state of the SDM photon statistics is always
super-poissonian, contrary to the conventional micromaser. On the
other hand, atom-atom correlations exhibit stronger nonclassical
features than in the micromaser dynamics, due to the stronger
entangling nature of the unitary atom-field interaction. We have
shown that, in the strong coupling regime, superpositions of
coherent states can be generated in the transient dynamics, whose
coherence vanishes towards its superpoissonian steady-state. Also,
we have shown the two-peaked structure of the SDM Pegg-Barnett
phase, explained by the underlying coherent gain mechanism. The
SDM model appears quite promising both as a nice theoretical tool,
with unusual access to exact analytical developments, and in view
of physical implementations, in cavity QED or in the optical
regime with high finesse Fabry-Perot cavities~\cite{m.l.}.

\vspace*{0.5cm}

\section{Acknowledgments}\label{sec6}
P. L. acknowledges financial support from the Bayerisches
Staatsministerium f\"ur Wissenschaft, Forschung und Kunst in the
frame of the Information Highway Project and E. S. from the EU
through the RESQ (Resources for Quantum Information) project.


\begin{thebibliography}{99}

\bibitem{m.m.} D. Meschede, H. Walther, and G. M\"uller,
\prl {\bf 54}, 551 (1985); G. Rempe, H. Walther, and
N. Klein, \prl {\bf 58}, 353 (1987).

\bibitem{qed} G. Raithel {\it et al.} in {\it Advances in
Atomic, Molecular and Optical Physics}, edited by P.R. Berman
(Academic, New York, 1994).

\bibitem{JC} E.T.Jaynes and F.W. Cummings, Proc.\ IEEE {\bf 51}, 89
(1963).

\bibitem{subp} G. Rempe, F. Schmidt-Kaler, and H. Walther,
\prl {\bf 64}, 2783 (1990).

\bibitem{TS} M. Weidinger, B. Varcoe, R. Heerlein, and H. Walther,
\prl {\bf 82}, 3795 (1999).

\bibitem{Fock} B. Varcoe, S. Brattke, M. Weidinger, and H.
Walther, \nat {\bf 403}, 743 (2000); S. Brattke, B. Varcoe, and
H. Walther, \prl {\bf 86}, 3534 (2001).

\bibitem{Englert1} B.-G. Englert, M. L\"off\/ler, O. Benson, B.
Varcoe, M. Weidinger, and H. Walther, Fortschr.\ Phys.\ {\bf 46},
897 (1998).

\bibitem{SAW} E. Solano, G. S. Agarwal, and H. Walther, \prl
{\bf 90}, 027903 (2003).

\bibitem{Haroche} J.M. Raimond, M. Brune, and S. Haroche,
\rmp {\bf 73}, 565 (2001).

\bibitem{CL} F. Casagrande, A. Lulli, and V. Santagostino,
\pra {\bf 65} 023809 (2002); F. Casagrande and A. Lulli,
J. Opt. B: Quantum Semiclass.\ Opt.\ {\bf 4} S260 (2002).

\bibitem{ion} E. Solano, R. L. de Matos Filho, and N. Zagury,
\prl {\bf 87} 060402 (2001).

\bibitem{CG} K. E. Cahill and R. J. Glauber, Phys.\ Rev.\ {\bf 177},
1857 (1969); ibid., 1882 (1969).

\bibitem{cats} M. Brune, E. Hagley, J. Dreyer, X. Maitre, A. Maali,
C. Wunderlich, J.M. Raimond, and S. Haroche,
\prl {\bf 77}, 4887 (1996).

\bibitem{SZ} See, for example, M. O. Scully and M. S. Zubairy,
\emph{Quantum optics} (Cambridge University Press, 1997).

\bibitem{Meystre} P. Filipowicz, J. Javanainen, and P. Meystre,
\pra {\bf34}, 3077 (1986); L.A. Lugiato, M.O. Scully, and
H. Walther, \pra {\bf 36}, 740 (1987).

\bibitem{Englert2} H.-J. Briegel, B.-G. Englert, N. Sterpi, and H.
Walther, \pra {\bf 49}, 2962 (1996).

\bibitem{QT} J. Dalibard, Y. Castin, and K. M{\o}lmer,
\prl {\bf 68}, 580 (1992).

\bibitem{CL2} For a recent application in the study of
the micromaser spectrum, see F. Casagrande, A. Ferraro, A. Lulli,
R. Bonifacio, E. Solano, and H. Walther,
\prl {\bf 90}, 183601 (2003).

\bibitem{PB} D.T. Pegg and S.M. Barnett, \pra {\bf 39}, 1665 (1989).

\bibitem{m.l.} K. An, J.J. Childs, R.R. Desari, and M. Feld, Phys.
Rev. Lett. {\bf 73}, 3375 (1994).

\end{thebibliography}
\end{document}